\documentclass{aa}  
\usepackage{graphicx}
\usepackage{txfonts}
\begin{document}
   \title{Accurate stellar masses in the multiple system T Tau}

   \author{G. Duch\^ene\inst{1}
          \and
          H. Beust\inst{1}
	  \and
	  F. Adjali\inst{1}
	  \and
	  Q. M. Konopacky\inst{2}
	  \and
	  A. M. Ghez\inst{2}
          }


   \institute{Laboratoire d'Astrophysique de Grenoble, BP 53, F-38041
              Grenoble cedex 9, France \\
              \email{gaspard.duchene@obs.ujf-grenoble.fr}
         \and
             Department of Physics and Astronomy, UCLA, Los Angeles,
              CA 90095-1562, USA}

   \date{Received ... ; accepted ...}

 
  \abstract
  {}
  {The goal of this study is to obtain accurate estimates for the
  individual masses of the components of the tight binary system
  T\,Tau\,S in order to settle the ongoing debate on the nature of
  T\,Tau\,Sa, a so-called infrared companion.}
  {We take advantage of the fact that T\,Tau\,S belongs to a triple
  system composed of two hierarchical orbits to simultaneously analyze
  the motion of T\,Tau\,Sb in the rest frames of T\,Tau\,Sa
  and T\,Tau\,N. With this method, it is possible to
  pinpoint the location of the center of mass of T\,Tau\,S and,
  thereby, to determine individual masses for T\,Tau\,Sa and
  T\,Tau\,Sb with no prior assumption about the mass/flux ratio of the
  system. This improvement over previous studies of the system
  results in much better constraints on orbital parameters.}
  {We find individual masses of 2.73$\pm$0.31$\,M_\odot$ for
  T\,Tau\,Sa and of 0.61$\pm$0.17$\,M_\odot$ for T\,Tau\,Sb (in
  agreement with its early-M spectral type), including the uncertainty
  on the distance to the system. These are among the most precise
  estimates of the mass of any Pre-Main Sequence star, a remarkable
  result since this is the first system in which individual masses of
  T\,Tauri stars can be determined from astrometry only. This
  model-independent analysis confirms that T\,Tau\,Sa is an
  intermediate-mass star, presumably a very young Herbig\,Ae
  star, that may possess an almost edge-on disk.}
  {}


  \maketitle


\section{Introduction}

The most robust method of determining the mass of stars is by
monitoring the Keplerian orbital motion of multiple systems. This
method is particularly valuable for contracting Pre-Main Sequence
(PMS) stars, such as T\,Tauri stars (TTS), for which mass estimates
from evolutionary models can differ by factors as large as 2 (e.g.,
Baraffe et al. 2002). However, there are only about 20 {\it
individual} TTS with well-determined masses, all of them having radial
velocity curves or a Keplerian circumstellar disk (see Hillenbrand \&
White 2004 for a recent review).

The triple system T\,Tauri is among the best studied PMS multiple
system (see, e.g, Beck et al. 2004). The optically bright T\,Tau\,N, a
$\sim2\,M_\odot$, $\sim1$\,Myr-old star, is associated with the
infrared(IR)-bright T\,Tau\,S tight binary system, which was resolved
less than a decade ago (Koresko 2000). The latter consists of
T\,Tau\,Sb, a deeply embedded, early M-type TTS, and of T\,Tau\,Sa, an
extremely red and highly variable object (Ghez et al. 1991) that has
been classified as an infrared companion (IRC). Despite two decades of
steady observations, the nature of that third component remains
largely unknown. It shares many spectral properties with heavily
embedded, Class\,I protostars, but its physical association with two
normal TTS seems to contradict this hypothesis if the system is coeval
(Koresko et al. 1997). Alternatively, Ghez et al. (1991) proposed that
T\,Tau\,Sa, which underwent a strong outburst, is a low-mass FU\,Ori
object. More recently, it has been suggested that T\,Tau\,Sa is an
intermediate-mass PMS star surrounded by an opaque edge-on
circumstellar disk (Duch\^ene et al. 2005).

Determining the mass of T\,Tau\,Sa from its orbital motion within this
triple system would help our understanding of its exact nature. While
it has been suggested that this system may be on an unbound orbit
(Loinard et al. 2003), other studies have found possible closed orbits
and derived a total mass of 2--6$\,M_\odot$ for T\,Tau\,S (Johnston et
al. 2003; Beck et al. 2004). Using a refined method, Johnston et
al. (2004a,b) have further attempted for the first time to determine
the {\it individual} masses of T\,Tau\,Sa and Sb. All of these studies
suggested that the IRC is the most massive component of the
system. However, they were based on simplifying (and sometimes
invalid) assumptions regarding the flux ratio and/or mass ratio of the
T\,Tau\,S system, resulting in additional sources of uncertainty.

Here, the combined orbital motion of the triple system is revisited
using an almost assumption-free, generic approach, in order to obtain a
precise estimate of the {\it individual} stellar masses within the
T\,Tau\,S tight system. This Letter is composed as follows: the
fitting method and the observations used in this analysis are
presented in Section\,\ref{sect:fit_data}, while the results and
implications are discussed in Section\,\ref{sect:results}.


\section{Approach and astrometric dataset}
\label{sect:fit_data}

Two types of astrometric data are available on the T\,Tau multiple
system. The first type is high-resolution IR data, which resolves all
three components since 1997. The second type is radio data, which has
detected for over two decades only two of the components, T\,Tau\,N
and a southern source. Though this southern source is shown by Loinard
et al. (2003) to be physically associated with T\,Tau\,Sb, Smith et
al. (2003) and Johnston et al. (2004a) argue that it might be offset
from it by up to 0\farcs01--0\farcs03. In our analysis, we assume that
the radio and IR source are indeed the same object. Only a much longer
IR follow-up of the system would allow one to determine whether this
assumption must be dropped.


\subsection{Fitting Method}
\label{sect:fitting}

Two general approaches have been used in previous attempts to fit the
orbital motion of the T Tau Sa-Sb tight binary system. First, it is
possible to consider in the fit only the IR datasets, which provide
the location of both components. In this case, however, the short
timespan probed since 1997 poorly constrains the orbit and a wide
range of total system masses is allowed (Beck et al. 2004). Including
our most recent IR measurements (see Section\,\ref{sect:data}) only
slightly improves this conclusion.

In a second approach, the separation between T\,Tau\,N and T\,Tau\,Sb,
estimated from the much longer radio observations, is transformed into
a T\,Tau\,Sa--Sb separation prior to fitting the orbital motion. In
this case, one has to make some assumptions regarding the unknown
location of T\,Tau\,Sa with respect to T\,Tau\,N. In the past, two
main categories of assumptions were made. First, some authors
considered that T\,Tau\,Sa follows a constant linear motion with
respect to T\,Tau\,N (Loinard et al. 2003; Johnston et al. 2003),
implicitly assuming that it is coincident with the center of mass of
T\,Tau\,S. However, while T\,Tau\,Sa is indeed more massive than its
companion, the mass of the latter is not negligible (see
Section\,\ref{sect:results}). Second, it was assumed that the
historical location of the photocenter of T\,Tau\,S, which has been
observed for over 15 years, can be used to trace the location of
T\,Tau\,Sa prior to 1997 (Johnston et al. 2004a,b). This requires one
to assume that the IR flux ratio within the T\,Tau\,S system remains
constant. Evidence for large variability in at least one of the two
components of the system is growing, however (Ghez et al. 1991; Beck
et al. 2004).

To limit as much as possible the number of assumptions in the
analysis, we decided to fit simultaneously two types of data, namely
the location of T\,Tau\,Sb in two rest frames: with respect to
T\,Tau\,Sa and to T\,Tau\,N. We did not use the separation between
T\,Tau\,N and T\,Tau\,Sa because it would not be independent from the
other two types of measurements.

\begin{table}
\caption{\label{tab:newdata} New astrometric and photometric data for
  the T Tau system.}
\begin{tabular}{cccc}
\hline
\hline
UT Date & Sep. ($\arcsec$) & P.A. (\degr) & $\Delta K$ \\
\hline
\multicolumn{4}{c}{T Tau Sa-Sb} \\
\hline
1997 Oct. 12 & 0.051$\pm$0.009 & 218$\pm$8 & 3.0$\pm$0.1 \\
2004 Dec. 19 & 0.116$\pm$0.004 & 294.1$\pm$1.4 & 0.76$\pm$0.12 \\
2005 Nov. 13 & 0.119$\pm$0.001 & 300.6$\pm$1.0 & 0.52$\pm$0.13 \\
\hline
\hline
\multicolumn{4}{c}{T Tau N-Sb} \\
\hline
1997 Oct. 12 & 0.733$\pm$0.009 & 181.1$\pm$0.6 & 5.6$\pm$0.2 \\
2004 Dec. 19 & 0.657$\pm$0.001 & 193.07$\pm$0.17 & 3.25$\pm$0.06 \\
2005 Nov. 13 & 0.650$\pm$0.001 & 193.81$\pm$0.15 & 2.42$\pm$0.12 \\
\hline
\end{tabular}
\end{table}

We assume that T\,Tau\,Sb follows a bound Keplerian orbit around
T\,Tau\,Sa and use the 141.5$\pm$2.8\,pc distance to the system
estimated by Loinard et al. (2005). We further assume that the motion
of the center of mass of T\,Tau\,S, which is a Keplerian orbit around
T\,Tau\,N, can be approximated by a parabolic motion over the 20
years of observations probed here. This is justified as the period of
this outer orbit is at least 400 years (Johnston et
al. 2004a,b). Fitting the motion of T\,Tau\,Sb in both rest frames
simultaneously allows us to determine individual masses from
astrometry only, as the ratio of the amplitude of the motion of
T\,Tau\,Sb in both frames is a direct measurement of the mass ratio.

We use a Levenberg-Marquardt $\chi^2$ minimization routine (Press et
al. 1992) that simultaneously fits for 14 independent parameters. The
T\,Tau\,Sa--Sb orbit is described by 7 parameters: semi-major axis
($a$), eccentricity ($e$), inclination with respect to the plane of
the sky ($i$), position angle of ascending node ($\Omega$, measured
Eastward from North), azimuth of periastron ($\omega$), time of
periastron passage ($t_P$) and orbital period ($P$). The
parabolic motion of the center of mass of T\,Tau\,S with respect to
T\,Tau\,N is described by 6 parameters defined by its right ascension
$\alpha_{{\mathrm S}} = \alpha_0 + v_\alpha (t-t_0) + a_\alpha
(t-t_0)^2$ and a corresponding equation for its declination, with
$t_0=2000.0$. Finally, the last parameter describes the mass ratio in
the system: $\mu=M_{\mathrm{Sb}}/M_\mathrm{S}$, where
$M_{\mathrm{Sb}}$ and $M_\mathrm{S}$ are the mass of T\,Tau\,Sb and of
the whole T\,Tau\,S system, respectively. Kepler's third law is used
to determine $M_\mathrm{S}$ from the values of $a$ and $P$; the
individual stellar masses $M_{\mathrm{Sa}}$ and $M_{\mathrm{Sb}}$ are
derived from the values of $M_\mathrm{S}$ and $\mu$. The minimization
routine was run with several tens of thousand initial guesses spanning
wide ranges for all fit parameters to ensure convergence on the
absolute minimum of the $\chi^2$ function.

The uncertainties on the fit parameters are estimated using a
bootstrap method, a robust method when performing a fit with many
(correlated) parameters. We created 10000 artificial datasets
containing as many points as the observed dataset, in which each point
is randomly drawn from a Gaussian distribution centered on the actual
measurement and whose width is given by the associated astrometric
uncertainty, and run the $\chi^2$ minimization routine for each
realization. The resulting distribution of values for each fit
parameter is characterized by a mean value equal to that of the best
fit and by a standard deviation that we consider as the estimate of
the uncertainty on the fit parameter. 

Because this fitting method is based on a fit of all available
measurements without further transformation, and because no parameter
is set to an ad hoc fixed value, it is very robust and essentially
assumption-free.


\subsection{IR astrometric dataset}
\label{sect:data}

A dozen high spatial resolution IR observations of the T\,Tau triple
system have been published; they usually provide simultaneously the
position of T\,Tau\,Sb with respect to both other stars. All
astrometric measurements listed in Beck et al. (2004) and Duch\^ene et
al. (2005) are included in our fit\footnote{The astrometric
measurements presented in Duch\^ene et al. (2002) suffered from an
incorrect astrometric calibration: the position angles used in this
study are 1.1\degr\ larger than was previously reported.}. Two
additional observations of the system were recently published by
Mayama et al. (2006). However, they are significantly offset
($\gtrsim3\sigma$) from all other quasi-contemporaneous measurements;
these measurements are likely biased or suffer from underestimated
uncertainties and they have been discarded from the analysis.

In addition to these measurements, we present new $K$-band speckle
interferometric datasets obtained on the 10m Keck I telescope in 1997,
2004 and 2005. For the last two epochs, the T\,Tau\,Sa--Sb system is
sufficiently well-resolved that the relative astrometry of all
components could be determined from shift-and-add images. On the other
hand, in 1997, the system was very tight ($\approx 1.25 \lambda/D$)
and traditional speckle analysis was conducted. Usual basic reduction
techniques were first applied to all datasets (see Ghez et al. 1993
for more details). Because T\,Tau is a hierarchical triple system, a
2-step method identical to that described in Duch\^ene et al. (2003)
was used to extract its properties. The new astrometric and
photometric data are presented in Table\,\ref{tab:newdata}.
 

\subsection{Radio astrometric dataset}

Loinard et al. (2003) and Johnston et al. (2003, 2004b) have published
a series of VLA centimeter observations of T\,Tau spanning 20 years,
from 1983 to 2003. Both groups independently analyzed the 1983, 1988,
1992, 1995 and 2001 datasets, and obtained very similar results. We
found that using either sets of results for these epochs yields very
similar fits\footnote{It is however incorrect to simultaneously fit
both, as was done by Tamazian (2004), since they come from the same
raw datasets.}. In the following, we use the 2\,cm astrometric results
from Johnston et al. (2003, 2004b) in order to ensure a maximum
uniformity among the radio datasets and their analysis, as well as the
1998 3.7\,cm measurement from Loinard et al. (2003), since no 2\,cm
observations were conducted at that epoch.


\section{Results and implications}
\label{sect:results}

Overall, we considered in our fit a total of 11 astrometric
measurements of the T\,Tau\,Sa--Sb system and 16 measurements of the
T\,Tau\,N--Sb system. This is the most extensive dataset fitted to
date on this system, considering that we did not try to fit the
historical location of the photocenter of T\,Tau\,S (but see
Section\,\ref{sect:masses}).


\subsection{Individual masses within T\,Tau\,S}
\label{sect:masses}

\begin{figure*}
\centering
\includegraphics[width=0.9\textwidth]{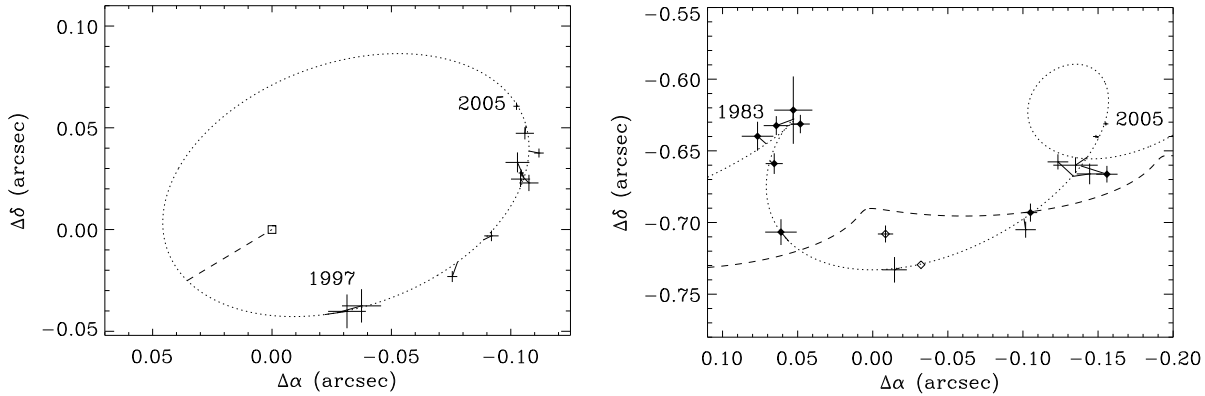}
\caption{ \label{fig:bestfit}Observed motion of T Tau Sb in the
  reference frame of T Tau Sa (left) and T Tau N (right). The dotted
  curves represent our best combined fit; each observation is
  connected to its modeled counterpart by a straight line. In the
  left-hand panel, the square indicates the fixed location of
  T\,Tau\,Sa and the dashed line connects it to the periastron. In the
  right-hand panel, filled diamonds represent the VLA measurements;
  the open diamonds indicate the 1998 VLA measurement excluded from
  the final fit and its fitted counterpart. In this panel, the dashed
  curve represents the motion of T Tau Sa.}
\end{figure*}

Fitting all measurements described in Section\,\ref{sect:fit_data}, a
best fit solution with $\chi^2_{\mathrm{red}}=1.88$ is obtained, a
slightly large value considering that there are 40 degrees of
freedom. In fact, the 1998 VLA measurement reported by Loinard et
al. (2003) is 4.1$\sigma$ away from the fitted value and contributes a
large amount to the total $\chi^2$. All other measurements are within
3$\sigma$ of the fit, and all but three less than 2$\sigma$ away from
it, as expected from a Gaussian distribution of the astrometric
uncertainties. This particular VLA measurement is the only one
included in our fit that was obtained at 3.7cm. Substantial
differences were noted in the past between 2cm and 3.7cm VLA
measurements for this source (Johnston et al. 2003, 2004b), and we
therefore consider this point as suspicious.

Excluding the 1998 VLA measurement, a best fit solution which is less
than 2.7$\sigma$ away from all remaining measurements and that has a
much more acceptable $\chi^2_{\mathrm{red}}=1.37$ is obtained with 38
degrees of freedom. No fit parameter is different by more than
1.4$\sigma$ from the previous solution, and we consider this new
solution as our best fit in the following. The best fit parameters are
listed in Table\,\ref{tab:fitpars} and the solution is illustrated in
Fig.\,\ref{fig:bestfit}. Note that it is not possible to distinguish
the ascending and descending nodes of the orbit from astrometry only:
an identical projected orbit could be described with
$\Omega=295.2$\degr\ and $\omega=-167.5$\degr.

\begin{table}
\caption{\label{tab:fitpars} Parameters of the best orbital solution.}
\begin{tabular}{cr@{\,$\pm$\,}l|cr@{\,$\pm$\,}l}
\hline
\hline
\multicolumn{6}{c}{Fit parameters}\\
\hline
$a$ ($10^{-3}$ \arcsec) & 82.1 & 1.8 & $\alpha_0$ ($10^{-3}$
\arcsec) & -20.7 & 4.0 \\
$e$ & 0.466 & 0.031 & $\delta_0$ ($10^{-3}$ \arcsec) &
-694.3 & 2.9 \\
$i$ (\degr) & 37.2 & 5.1 & $v_\alpha$ ($10^{-3}$ \arcsec/yr) &
-9.12 & 0.54 \\
$\Omega$ (\degr) & 115.2 & 6.8 & $v_\delta$ ($10^{-3}$ \arcsec/yr) &
1.71 & 0.39 \\
$\omega$ (\degr) & 12.5 & 10.3 & $a_\alpha$ ($10^{-4}$
\arcsec/yr$^2$) & 1.09 & 0.50 \\
$t_P$ & 1995.86 & 0.21 & $a_\delta$ ($10^{-4}$ \arcsec/yr$^2$) &
1.18 & 0.58 \\
$P$ (yr) & 21.66 & 0.93 & $\mu$ & 0.183 & 0.044 \\
\hline
\hline
\multicolumn{6}{c}{Derived quantities$^{\mathrm{a}}$}\\
\hline
$A$ (AU) & 11.59 & 0.35 & $M_{\mathrm{S}}$ ($M_\odot$) &
3.34 & 0.36 \\
$M_{\mathrm{Sa}}$ ($M_\odot$) & 2.73 & 0.31 & $M_{\mathrm{Sb}}$
 ($M_\odot$) & 0.61 & 0.17 \\
\hline
\end{tabular}
\begin{list}{}{}
\item[$^{\mathrm{a}}$] The uncertainty on the derived quantities
  include that induced by the distance estimate.
\end{list}
\end{table}

We then performed a consistency check regarding the historical
measurements of the IR photocenter of the T\,Tau\,S binary. While
strong variability prevents using these measurements in the fit, it
must be verified that the photocenter is located along the line
joining the fitted location of T\,Tau\,Sa and T\,Tau\,Sb at all
epochs. Satisfyingly, the photocenter lies within 2.4$\sigma$ of the
predicted line connecting the two components for all measurements
listed in Ghez et al. (1995) and Roddier et al. (2000), giving more
confidence in the validity of the fit.

Including the uncertainty induced by the distance estimate, the total
relative uncertainty on $M_{\mathrm{Sa}}$ is about 11\%, among the
most accurate individual mass estimates for a TTS that does not belong
to an eclipsing binary system (Hillenbrand \& White 2004). The
uncertainty on $M_{\mathrm{Sb}}$ is about 28\%, still a remarkably
accurate estimate for a TTS, considering that only astrometric
measurements were taken into account in this analysis.


\subsection{Discussion}

The results obtained here are in good agreement with those presented
by Johnston et al. (2004a,b), though with uncertainties that are at
least twice as small for all parameters: the best estimate for
$M_{\mathrm{Sa}}$ to date was 2.1$\pm$0.8$\,M_\odot$, for instance
(Johnston et al. 2004b). This significant improvement is due to a
combination of factors: the increased time coverage of high-resolution
IR datasets, a simpler characterization of the T\,Tau\,N--S orbit
(parabolic motion instead of a Keplerian orbit with several parameters
fixed to ad hoc values), the improved distance estimate obtained by
Loinard et al. (2005), and the smaller number of assumptions used in
our analysis. In this work, the only assumptions that could be
disputed are the parabolic approximation of the orbital motion of
T\,Tau\,S around T\,Tau\,N over 20 years and the match between the
radio and IR location of T\,Tau\,Sb. The fact that a very satisfying
fit was obtained appears to validate these hypotheses.

The mass derived for T\,Tau\,Sb is in excellent agreement with its
known, early-M spectral type (Duch\^ene et al. 2002), as already
discussed by Johnston et al. (2004a,b). For the first time, however,
an accurate estimate of the mass of T\,Tau\,Sa is obtained, leaving
little doubt that it is somewhat more massive than T\,Tau\,N itself
and suggesting that T\,Tau\,Sa is a young embedded Herbig\,Ae
star. This is the first time that this fact is established with such
certainty for an IRC. As discussed in Duch\^ene et al. (2005), it is
likely that T\,Tau\,Sa is surrounded by a compact and partially opaque
circumstellar disk that accounts for the apparent faintness of this
intrinsically bright object. Given the orbital parameters of the
T\,Tau\,Sa--Sb system, the radius of this disk must be
$\lesssim3$\,AU, i.e., half the periastron distance. We remark that
this disk must be very inclined with respect to the orbital plane
($\Delta i \gtrsim 45$\degr), raising the possibility of gravitational
instabilities.

Within 10 years or so, spatially-resolved IR observations of the
T\,Tau triple system will allow the derivation of accurate orbital
parameters for T\,Tau\,S without making use of the radio
datasets. With such an analysis, it will be possible to determine
whether the southern radio component of T\,Tau is indeed arising from
the chromosphere of T\,Tau\,Sb or whether it is a separate object that
is simply associated with this object.

On the shorter term, the dynamical stability of the T\,Tau system must
be studied in more detail: the system contains three stars within
100\,AU of each other, each of them possessing a circumstellar
disk. The overall velocity of T\,Tau\,S with respect to T\,Tau\,N is
on order 6\,km/s, very similar to the velocity of a 100\,AU-radius
circular orbit for a 5\,$M_\odot$ system (assuming
$M_\mathrm{N}=2\,M_\odot$). This suggests that the outer orbit is seen
close to face-on and has a low eccentricity. If this is indeed
the case, the application of the stability criterion proposed by
Eggleton \& Kiseleva (1995) suggests that the system could be
marginally stable. Moreover, the stability of the circumstellar disks
is not necessarily ensured in such a tight system. This motivates a
follow-up dynamical study similar to that performed for the quadruple
system GG\,Tau (Beust \& Dutrey 2005, 2006) to determine the possible
solutions for the outer orbit that preserve the three components and
their disks for at least 1\,Myr, i.e., the age of the system. A
preliminary analysis, which will be presented in a forthcoming paper,
suggests that the three-body system is stable over a few Myrs, whereas
the fate of the circumstellar disks is not yet established.


\begin{acknowledgements}
      We would like to acknowledge our referee, Mike Simon, for his
      insightful remarks. We thank J. Lu for her help in calibrating
      the NIRC camera and L. Saug\'e and X. Delfosse for clarifying
      discussions on the methods to derive uncertainties. We also
      thank L. Loinard for providing us with his published radio
      measurements. This work was funded in part by the NASA
      Astrobiology Institute and the NSF Science \& Technology Center
      for AO, managed by UCSC (AST-9876783). The data presented herein
      were obtained at the W.M. Keck Observatory, which is operated as
      a scientific partnership among the California Institute of
      Technology, the University of California and the National
      Aeronautics and Space Administration. The Observatory was made
      possible by the generous financial support of the W.M. Keck
      Foundation. The authors wish to recognize and acknowledge the
      very significant cultural role and reverence that the summit of
      Mauna Kea has always had within the indigenous Hawaiian
      community. We are most fortunate to have the opportunity to
      conduct observations from this mountain.
\end{acknowledgements}


\end{document}